\documentclass[11pt]{article}

\def\fnote#1#2{\begingroup\def\thefootnote{#1}\footnote{#2}\addtocounter{footnote}{-1}\endgroup}

\usepackage{hyperref}
\hypersetup{
    colorlinks=true, linkcolor=blue, filecolor=black, urlcolor=blue, 
    pdftitle={Overleaf Example}
    pdfpagemode=FullScreen,
    }
\usepackage{url}

\usepackage{amssymb}
\usepackage{makeidx}
\usepackage{epsf}
\usepackage{graphicx}        % standard LaTeX graphics tool
\usepackage{cite}

\makeindex

\def\inbar{\vrule height1.5ex width.4pt depth0pt}
\def\IB{\relax{\rm I\kern-.18em B}}
\def\IC{\relax\,\hbox{$\inbar\kern-.3em{\rm C}$}}
\def\ID{\relax{\rm I\kern-.18em D}}
\def\IE{\relax{\rm I\kern-.18em E}}
\def\IF{\relax{\rm I\kern-.18em F}}
\def\IG{\relax\,\hbox{$\inbar\kern-.3em{\rm G}$}}
\def\IH{\relax{\rm I\kern-.18em H}}
\def\II{\relax{\rm I\kern-.18em I}}
\def\IK{\relax{\rm I\kern-.18em K}}
\def\IL{\relax{\rm I\kern-.18em L}}
\def\IM{\relax{\rm I\kern-.18em M}}
\def\IN{\relax{\rm I\kern-.18em N}}
\def\IO{\relax\,\hbox{$\inbar\kern-.3em{\rm O}$}}
\def\IP{\relax{\rm I\kern-.18em P}}
\def\IQ{\relax\,\hbox{$\inbar\kern-.3em{\rm Q}$}}
\def\IR{\relax{\rm I\kern-.18em R}}
\def\IT{\relax{\rm I\kern-.18em T}}
\def\ZZ{\relax{\sf Z\kern-.4em Z}}

\def\nablaslash{\relax{\rm /\kern-.28em \nabla}}

\def\a{\alpha}         
\def\e{\epsilon}      
\def\L{\Lambda}     \def\si{\sigma}

\def\cA{{\cal A}}

   \def\cR{{\cal R}}

 \def\oD3{{\overline \rmD 3}}

\def\fnote#1#2{\begingroup\def\thefootnote{#1}\footnote{#2}\addtocounter
{footnote}{-1}\endgroup}
\def\beq{\begin{equation}}
\def\eeq{\end{equation}}
\def\bea{\begin{eqnarray}}
\def\eea{\end{eqnarray}}

\def\lleq#1{\label{#1}\eeq}

\def\notin{\ \hbox{{$\in$}\kern-.51em\hbox{/}}}

\def\lra{\longrightarrow}

  \def\E1Fq{E_1/\IF_q}

\def\rmD{{\rm D}}

\def\rmII{{\rm II}}  \newcommand{\rmIII}{{\rm III}}

 \def\rmD{{\rm D}}

    \newcommand{\rmpiv}{{\rm piv}}

     \def\rmrh{{\rm rh}}

        \def\rmPl{{\rm Pl}}

\def\notdiv{{\relax{~|\kern-.35em /~}}}

\def\boxit#1{
\vbox{\hrule height1pt\hbox{\vrule width1pt\kern0.3cm
\vbox{\kern0.3cm\hbox{$\displaystyle#1$}\kern0.3cm}\kern0.3cm\vrule
width1pt}\hrule height1pt}}

\begin{document}
\parindent=0pt

%\phantom{what} \hfill \today

\baselineskip=18pt
\parskip=0.1truein
\parindent=0pt

\hfill \phantom{ {\bf Draft}} \break
\phantom{what} \hfill  \phantom{\today}~\break

\vskip 0.7truein

\centerline{\LARGE {\bf ACT Implications for  Hilltop Inflation}}

\vskip .4truein

\centerline{\sc Monika Lynker\fnote{1}{mlynker@iu.edu} and Rolf Schimmrigk\fnote{2}{rschimmr@iu.edu, netahu@yahoo.com}}

\vskip .2truein

\centerline{Dept. of Physics}

\centerline{Indiana University South Bend}

\centerline{1700 Mishawaka Ave., South Bend, IN 46615}

\vskip 1.4truein

\centerline{\bf Abstract}

\begin{quote}
We analyze in a systematic way the implications for the classes of hilltop and hilltop-squared inflation of the recent DR6 observations 
by the Atacama Cosmology Telescope collaboration. We find that the reported shift in the spectral index leads to parameter ranges 
for these models that are significantly reduced when compared to the results obtained from the {\sc Planck} observations. We mainly focus on the 
more dramatic implications for the hilltop-squared class, but along the way we  also highlight the milder impact on  the class of hilltop models.
\end{quote}

\renewcommand\thepage{}
\newpage
\parindent=0pt

\pagenumbering{arabic}

\vfill \eject

\baselineskip=16pt
\parskip=0.01truein
\parindent=0pt

\tableofcontents

\vskip .6truein
\baselineskip=20.1pt
\parskip=0.15truein
\parindent=0pt

\section{Introduction}

The high-desert Atacama Cosmology Telescope (ACT) collaboration has in its recent data release \cite{act25a, act25b} 
reported a scalar spectral index that has shifted away from the {\sc Planck 2018} result \cite{planck18a-aghanim, planck18b-akrami, planck19-aghanim} to a higher value. 
Combining the ACT observations with those of the BICEP/Keck collaboration \cite{bk18-a21etal},  and the DESI data \cite{desi24a, desi24b, desi25, desi25b} 
leads to a spectral index reported in \cite{act25a} as  $n_{\cR\cR} =0.974\pm 0.003$. 
This is a significant change that raises the question what precisely the implications are of this new measurement for inflationary models, 
leading to a campaign that has recently been taken up in several papers, including
 \cite{klr25, aoy25, b25etal, dir25, s25, k25etal, dk25, rs25, ggyz25,  dx25, zsv25, hpp25,  gkrr25, lyg25, y25, gkt25, bcl25, y25etal, 
         aak25, pcl25, mmc25, fk25, hm25, pallis25a, oo25, w25, m25, c25etal, os25, ar25, g25etal, p25, hls25, wb25, pallis25b, 
         pal25, m25etal, f25etal, r25etal, ks25}. 
         
 Motivated by the results of the ACT collaboration we revisit in this paper the classes of hilltop and hilltop-squared models 
 of inflation to analyze the changes that are imposed on these theories. Individual models in this class have received
 much attention over the past three decades,  see for example refs. \cite{km95, bl05, g20, hs21,  lr22} where many more references
 can be found.  These two classes were systematically analyzed in depth with 
the {\sc Planck} constraints some time ago \cite{ls23}, in particular the behavior of these models
 with respect to the variation of the parameter $\mu$ associated to the inflaton itself. This leads 
 to scaling relations for its observables with characteristic exponents that provide a classification scheme 
 that extends also to other classes of models, such as multifield modular inflation models \cite{ls22}. 
It was shown there that for low exponents $p$ these models do not admit sub-Planckian $\mu$
while for larger $p$ the sub-Planckian regime becomes accessible. 
It was also noted in ref. \cite{ls23} that hilltop and hilltop-squared models 
at low values of $p$ would come under increased pressure if the constraints 
on the spectral index were to change even without any further changes to the tensor ratio. 
It is therefore of interest to investigate what the impact is of the ACT result on these classes of models.
In the present paper we give a detailed analysis of the class of hilltop-squared models because they turn out to be more severely constrained 
by the ACT result, as compared to the {\sc Planck} data. A similar analysis can be done by considering the SPT data \cite{spt25-c25etal}, 
but our focus in the present analysis is on the ACT induced results. Along the way we however also describe the milder implications 
of the ACT spectral shift for the hilltop models and we compare these two classes of theories.

This paper is organized as follows.
In section 2 we briefly describe the structure of the class of inflation models in different approximation schemes. 
In section 3 we report on the significant impact of the ACT result on these models, and in 
section 4 we summarize our findings.

\vskip .3truein

\section{Hilltop and hilltop-squared models}

Different models among the classes of hilltop (HI) and hilltop-squared models (HSI) have been investigated in a large number of papers,
with early references  \cite{km95, bl05} and continuing more recently in refs. \cite{g20, hs21,  lr22, ls23}, where many more references 
can be found.  The models considered in the present paper are part of a two-parameter family of  theories described by potentials of the form
\beq
 V ~=~ \L^4 \left( 1- \left(\frac{\phi}{\mu}\right)^p \right)^n,
 \lleq{hsi-potential}
 with the overall energy scale $\L$ and the inflaton parameter $\mu$. The former is determined by the CMB power spectrum, 
 while $\mu$ is in principle a free parameter, which however is constrained by the CMB observables and the number of e-folds.
 The two classes of models relevant here are hilltop inflation ($n=1$) and hilltop-squared inflation ($n=2$). 
 While these models show a  similar behavior when constrained by the {\sc Planck} constraints,   the advantage of hilltop-squared models, 
 as compared to the hilltop models, is that the potential is bounded from below, leading to the standard picture for the reheating regime 
 via an oscillating inflaton. 
  In the present section we briefly describe the key variables of these classes of models.
 
 The observables considered in the present paper include the spectral index $n_{\cR\cR}$ 
 of the Lukash-Bardeen perturbation \cite{l80, b80, w08}
 \beq
  \cR ~=~ H\delta u ~-~ \psi,
  \eeq
 which in the slow-roll approximation can be obtained via the slow-roll parameters $\e_V$ and $\eta_V$, defined as 
 \beq
 \e_V ~=~ \frac{1}{2}M_\rmPl^2 \left(\frac{V'}{V}\right)^2, ~~~~\eta_V ~=~ M_\rmPl^2 \frac{V''}{V}.
 \eeq
 These two parameters are related as
 \beq
  \eta_V ~=~ \left[ 2 \left(1- \frac{1}{np} \right) - \frac{2}{n} \left(1-\frac{1}{p}\right)   \left(\frac{\mu}{\phi}\right)^p\right] \e_V,
  \lleq{etaV-vs-eV}
  a relation that will be useful later in this paper. 
  
  The spectral index is given in the slow-roll approximation by 
  \beq
 n_{\cR\cR}(\phi) ~=~ 1 - \frac{np\frac{M_\rmPl^2}{\mu^2}}{\left(1-\left(\frac{\phi}{\mu}\right)^p\right)^2}\left(\frac{\phi}{\mu}\right)^{p-2}
    \left[  2(p-1)   + (np+2)  \left(\frac{\phi}{\mu} \right)^p  \right],
 \eeq
while the tensor-to-scalar ratio takes the form
 \beq
 r(\phi) ~=~  8n^2p^2    \frac{M_\rmPl^2}{\mu^2} 
  \frac{ \left(\frac{\phi}{\mu}\right)^{2(p-1)} }{\left(1-\left(\frac{\phi}{\mu}\right)^p\right)^2},
 \eeq
 which is four times the tensor ratio in hilltop inflation. 
The number of e-folds $N_*$  is given by 
 \beq
  N_* ~=~ \frac{\mu^2}{npM_\rmPl^2} 
    \left[ \frac{1}{p-2}  \left( \left(\frac{\mu}{\phi_*}\right)^{p-2}  ~-~ \left( \frac{\mu}{\phi_e}\right)^{p-2} \right) 
         - \frac{1}{2} \left( \left(\frac{\phi_e}{\mu}\right)^2 -  \left(\frac{\phi_*}{\mu}\right)^2\right) \right],
  \lleq{sr-efolds}
hence for general $n$ the higher exponent e-folds are suppressed by a factor of $1/n$.

In order to get analytical scaling functions of $N$ for these observables it is necessary to make further assumptions that 
go beyond the slow-roll approximation and restrict the inflaton relative to $\mu$ in these models. First, in
 the small field approximation in which this constraint 
applies during the whole period of inflation the spectral index can be expressed as
\beq
 n_{\cR\cR}^\rmII(N) 
  ~=~   1 -  \frac{ 2np(p-1)}{np(p-2) N +  \left(\frac{n^2p^2}{2}\right)^{(p-2)/(2(p-1)} \left(\frac{\mu}{M_\rmPl}\right)^{p/(p-1)} }
  \lleq{nRR-SFII}
  and
  \beq
   r_*^\rmII(N) ~=~
   \frac{ 8n^2p^2  \left(\frac{\mu}{M_\rmPl}\right)^{2p/(p-2)} }{ 
   \left[ np(p-2) N + \left(\frac{n^2p^2}{2}\right)^{(p-2)/(p-1)}  \left(\frac{\mu}{M_\rmPl}\right)^{p/(p-1)}\right]^{2(p-1)/(p-2)}}
  \lleq{r-SFII}
  It is manifest from these relation that the small field approximation per se still retains information about the energy parameter $\mu$, as
  well as the model type parameters $p$ and $n$.
  
If it is  however further assumed that the parameter $\mu$ is much smaller than the Planck mass the functional relation leads 
to a simple scaling relation of the form
\beq
 n_{\cR\cR}^\rmIII ~=~ 1 - \frac{2(p-1)}{(p-2)N}.
 \lleq{nRR-SFIII}
 This relation is not only not sensitive to variation of  $\mu$ but also independent of the class of models,
defined by the exponent $n$. It thus represents a dramatic degeneration of the spectral index over the parameter spaces of an 
infinite number of models.  
 In the same SF.III approximation the $\mu$-dependence of the tensor ratio on the other hand reduces to an overall factor
 \beq 
 r^\rmIII ~=~  \left(\frac{\mu}{M_\rmPl}\right)^{2p/(p-2)}  \frac{8n^2 p^2}{[np(p-2) N]^{2(p-1)/(p-2)}}.
 \lleq{r-SFIII}
 Similar scaling laws of the type III considered here for hilltop models have received much attention  in the literature because of 
their simplicity and universality, in particular in the context of the models of Starobinsky and Higgs inflation. In combination with 
the canonical constraints on the number of e-folds these relations provide an easy comparison of this regime to the data provided by 
CMB and galaxy survey collaborations. To answer the question whether a model is viable or not however needs the inclusion of the 
   $\mu$-dependence of the observables that is present in all inflationary models with nontrivial potentials. 
   We will illustrate this  later in this paper when we compare our hilltop  models with the data. 
   In particular the spectral index considered in the regime III, i.e. in the small field approximation and for sub-Planckian 
   parameter values $\mu$,  leads to values that are dramatically different from the values that the models can reach once these
   assumptions are dropped.   The spectral index in the limit $n_{\cR\cR}^\rmIII$ would for instance lead to the conclusion that all hilltop and hilltop-squared 
  models are excluded by the ACT results. We will show below that this however is not the case for these models. 
 While in the {\sc Planck} regime for the observables there exist regions in field and parameter space in which even the regime III is
 viable,  the new observations exclude these parameter regions for both  hilltop-squared and hilltop models
  and hence this small field and small $\mu$ is not viable in either class. This however merely means that the full parameter space 
  of these models has not been taken into account.
  
   It is in this context of interest to note that the hilltop spectral index in the regime  $n_{\cR\cR}^\rmIII$ has a limiting form for large $p$ that is identical to the form
   of a universal index that is often considered for other  models:
  \beq
   n_{\cR\cR}^\rmIII ~~\stackrel{p~{\rm large}}{\lra} ~~~ n_{\cR\cR} ~=~ 1 ~-~ \frac{2}{N}.
   \eeq
 This limiting form of the spectral index of the class of hilltop models arises for example in the class of Starobinsky type models that generalize 
 the original model \cite{s80, w84, m87} by  
 including a parameter $\mu$ in the same normalizing way as the hilltop models.  In this case the spectral index again retains the 
 dependence on $\mu$ in the slow-roll approximation, as does the tensor-to-scalar ratio. The inversion of the e-fold formula is again 
 only possible in a particular limit of the inflaton range, in this case the large field limit. The resulting function $n_{\cR\cR}(N)$ still 
 depends on $\mu$ and the form above can be obtained by further assuming that $\mu$ is in the sub-Planck regime.
 This shows that comparing such limiting forms of the spectral index with the CMB data only tests inflationary models in 
 particular regimes of both the inflaton and the energy parameter $\mu$. In order to see whether a model can accommodate the 
 data it is therefore of interest to keep the parameter dependence of the CMB observables. Similar remarks apply to other 
 models with nontrivial potentials because these always involve $\mu$-parameters.

 Including reheating in these models leads to an additional constraint via the number of reheating e-folds $N_\rmrh$ that can 
 be written in terms of the dimensionless potential $f = V/\L^4$ as
 \beq
 N_\rmrh ~=~ \frac{4}{1-3w_\rmrh} \left(-N_*  ~+~ \frac{1}{4} \ln \frac{(f'_*)^2}{f_*f_e} ~+~ \frac{1}{2} \ln \frac{M_\rmPl}{\mu} 
 ~+~ \si_\rmpiv \right),
 \lleq{Nrh-for-general-f}
 where $f_*$ is the dimensionless potential evaluated at $\phi_*$, $f_e$  is its value at the end-of-inflation, and
  $f'$ is the dimensionless derivative of $f$. The parameter  $\si_\rmpiv$ is a model independent constant, defined as
 \beq
  \si_\rmpiv ~=~ \ln \frac{a_0T_0}{k_*} ~-~ \frac{1}{3}\ln \frac{11g_\rmrh}{43} - \frac{1}{4} \ln \frac{405}{\pi^2 g_\rmrh} 
    + \frac{1}{4} \ln (12\pi^2 \cA_\cR),
  \eeq
  where $\cA_\cR$ is the CMB amplitude, $k_*$ is the pivot scale, $g_\rmrh$ counts the degrees of freedom, and $T_0$ is the 
  temperature. More details   about reheating in hilltop models can be found in \cite{ls23}.
  
It was shown in \cite{ls23} that the behavior of hilltop-squared models ($n=2$) {\bf is  similar} to the behavior of the hilltop models 
($n=1$) if the constraints of the {\sc Planck} satellite are imposed.  In the remainder of this paper we will analyze the 
implications of the constraints obtained from the Atacama Cosmology Telescope, in combination with other data sets. 
It turns out that while for both classes of models the new measurement leads to significant changes in the viable 
parameter space, the impact is more dramatic  for the hilltop-squared models, with milder implications for the hilltop models.
We highlight the differences between these two classes of models in light of the ACT result.

\vskip .2truein

\section{P-ACT-LB-BK18 DR6 constraints on hilltop models}

 The data release DR6  of the Atacama Cosmology Telescope collaboration \cite{act25a, act25b} provides a new fit 
 to a combined data set that includes besides the ACT data  the {\sc Planck} data \cite{planck18a-aghanim, planck18b-akrami, planck19-aghanim}, the
 CMB lensing data from ACT and {\sc Planck}, as well as the BK18 collaboration results \cite{bk18-a21etal}  and the
  DESI BAO data \cite{desi24a, desi24b, desi25, desi25b}.  
 The combination of these data sets is referred to by the ACT collaboration as P-ACT-LB-BK18 and  their new constraints 
  are summarized in \cite{act25b} in a graphical compilation that we recall here in Fig. 1 for reference.
  \parskip=0pt
\begin{center}
\includegraphics[scale=0.5]{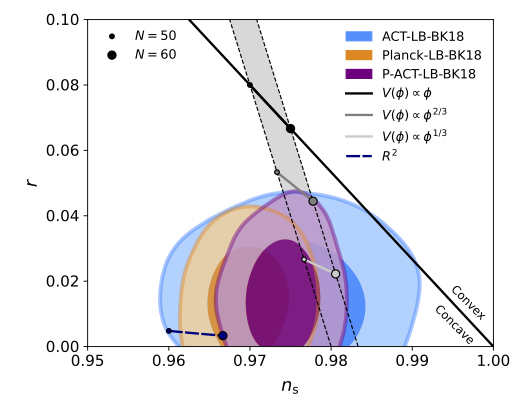}
\end{center}
\begin{quote}
{\bf Fig. 1} ~Results for the spectral index and the tensor-to-scalar ratio from the Atacama Cosmology Telescope, 
in combination with other data sets. Included here are several models that have been analyzed by 
the collaboration \cite{act25b}.  We will present the hilltop-squared results in this plane in Fig. 2 below
and a hilltop model in Fig. 5.
\end{quote}

\parskip=0.1truein

It was mentioned in the introduction already that the results of ref. \cite{ls23} show 
that hilltop and hilltop-squared models at low values of $p$ would come under increased pressure if the constraints 
on the spectral index were to change even without any further changes to the tensor ratio because this would further restrict the range of 
the parameter $\mu$ that enters the potential (\ref{hsi-potential}).  It is therefore of interest to investigate the implications of the new 
constraints for this class of models. To make the changes transparent and provide a better perspective 
on the new data it is useful to present the results obtained in the present paper in tandem with the {\sc Planck} based results 
reported  in \cite{ls23}.
 There the spectral index was constrained to lie in the interval $n_{\cR\cR} = 0.965 \pm 0.005$,  while the 
tensor ratio was bounded as $r\leq 0.035$, motivated by the {\sc Planck} satellite constraints. The number of e-folds was restricted 
to lie in the interval $N_* \in [50, 70]$. Keeping the latter two as before,
 we show in Fig. 2 that these results are affected in a rather significant way by shift  of the spectral index reported in ref. \cite{act25a, act25b}. 
 \begin{center}
 \includegraphics[scale=0.62]{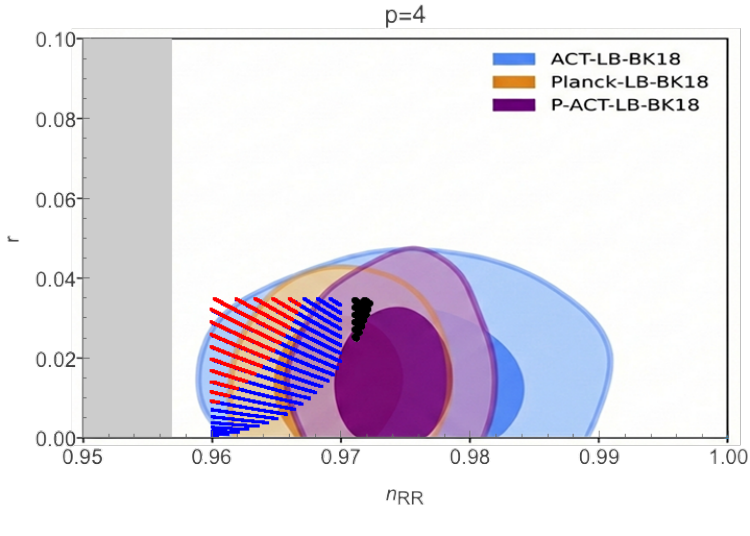}
 \includegraphics[scale=0.62]{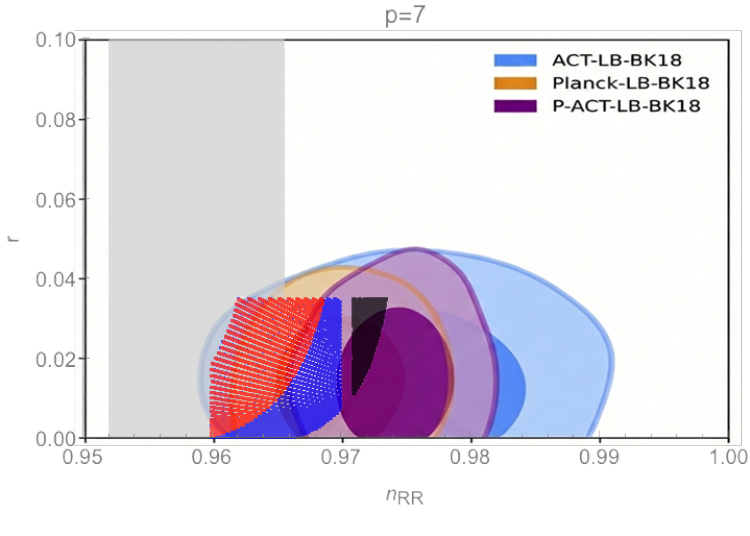}
 \includegraphics[scale=0.62]{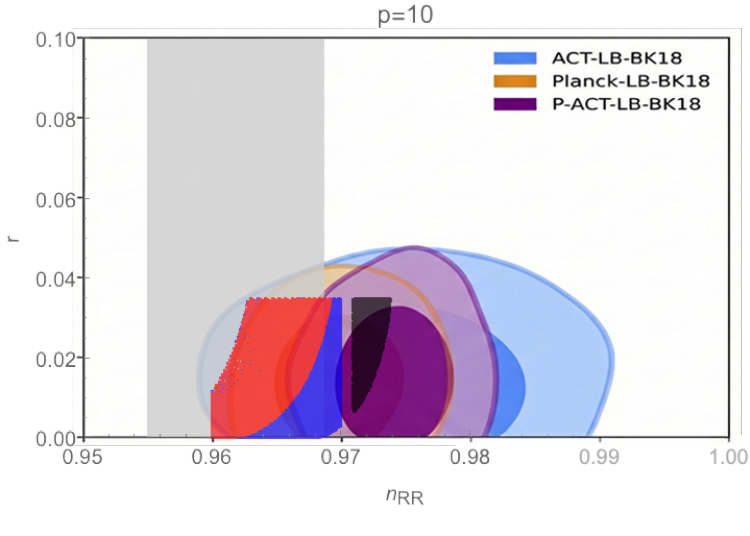}
 \includegraphics[scale=0.62]{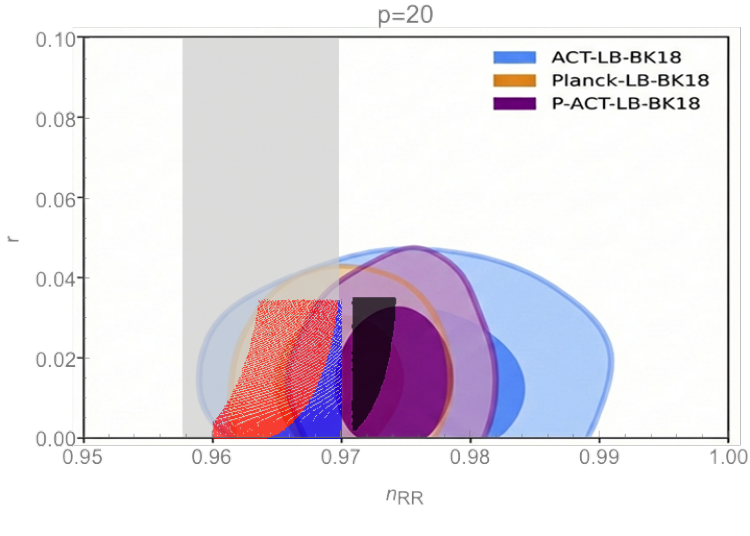}
 \end{center}
\begin{quote}
{\bf Fig. 2}~ The four panels show the results for the tensor ratio along the vertical axis against the spectral index 
 along the horizontal axis induced by both the  {\sc Planck} data (blue) and by the Atacama results (black) for 
 hilltop-squared models with $p=4, 7, 10, 20$, arranged from the 
top left to the right bottom panels.  The red region indicates those models for which matter dominated reheating is viable.
The vertical bands show the range of the spectral index in the regime of small fields and low values of the parameter $\mu$, 
as given in eq. (\ref{nRR-SFIII}),
for  the e-fold range $N_* \in [50, 70]$. This shows that while the regime of  small field and small $\mu$ HSI is 
disfavored, there are parameter regions in which hilltop-squared inflation is compatible with the 
the P-ACT-LB-BK18 data. See the text for more discussion.
\end{quote}

These results have a number of implications which we now highlight.
First, it is of importance that the regime of small field inflation and small sub-Planckian parameter values $\mu\ll M_\rmPl$ for the 
$p=4$ model of the hilltop-squared class is disfavored because the resulting SF.III spectral index, given in eq. (\ref{nRR-SFIII}) 
and indicated by the vertical band in the $p=4$ panel, is outside the confidence regions of all the combinations of data sets 
considered by the ACT collaboration \cite{act25b}, since in this case the spectral index is constrained as
\beq 
 n_{\cR\cR}^\rmIII ~=~ 1 - \frac{3}{N_*} ~\in ~[0.940, 0.957].
 \eeq
The fact that the  regime of small $\phi$ and  sub-Planckian $\mu$ is disfavored for the $p=4$ model at more than $2\si$ raises the question whether 
 this model can provide a better fit to the data when these constraints on the inflaton field and the parameter $\mu$ are not imposed. 
Our results in the $p=4$ panel of Fig. 2 show that by admitting super-Planckian parameter values $\mu$, models can be obtained within the 
2$\si$ region of the P-ACT-LB-BK18 analysis, with the $1\si$ region reached only with the upper limit of the $\mu$ values. This upper bound of $\mu$ 
 is determined by the bound on the tensor ratio $r$.
 
 The fact that the low$-p$ hilltop-squared models have come under increased pressure from the recent data provides a further motivation 
 to investigate how models with larger $p$ compare to this data. Such models have previously been investigated with a different focus in 
 ref. \cite{ls23} because they present natural models to consider within the framework of hilltop-squared inflationary theory and there is no
  reason to exclude them. In Fig. 2 we extend our results for $p=4$ to  a representative selection of models at $p=7, 10, 20$, which illustrate the behavior of 
  this class of models. These results show that for $p>4$ the spectral index again reaches the 
 Atacama regime and that  the upper bound of the spectral index leads to a completion of these diagram 
 in that none of the considered models actually covers the full domain that is allowed by \cite{act25a, act25b}.
 Only the models with higher $p$ reach the value $n_{\cR\cR} =0.974$ and as $p$ increases the SF.III regime ranges do eventually overlap with the 
 $2\si$ likelihood region of P-ACT-LB-BK18.  It also shows that for none of the models considered 
 here matter dominated reheating is possible. These results establish a pattern for all the models 
that we have checked for values of $p$ up to one hundred.

A feature of the hilltop-squared models that is induced by the variation of the parameter $\mu$ is  the stratification that can be seen 
 in Fig. 2. This parameter  remains undetermined by the scalar power spectrum amplitude but is constrained by the bounds on the 
 spectral index and the tensor-to-scalar ration.
The variable $\mu$ is varied in the scans in steps of one Planck mass and the cut-off at the upper end of the distributions is imposed by the Planck bound on 
the tensor ratio, which leads to an upper bound for this parameter. The graphs show that the lower bound of $\mu$ consistently shifts higher for the P-ACT-LB-BK18 
result in comparison to the {\sc Planck} data. The upper bound of $\mu$ increases as well, hence the whole viable domain of the 
parameter space shifts to higher $\mu$. As a result $\mu$ never reach the sub-Planckian regime for any of the models,
making the behavior of the models compatible with the ACT data fundamentally different from the behavior of the models 
compatible with the {\sc Planck} regime.

The $\mu$-stratifications that appear in Fig. 2 show that in hilltop-squared inflation the tensor-to-scalar ratio $r$ follows scaling relations in dependence 
of the spectral index, with scaling parameters that depend on the parameter $\mu$ associated to the inflaton.
This was analyzed in depth quantitatively in \cite{ls23},  where these scaling parameters were recognized  as useful characteristics of inflationary models, 
following the earlier observation \cite{ls22} of such scaling behavior in modular inflation models at level one \cite{rs14, rs16, rs21}, as well as for 
higher modular level \cite{ls19}. Taking this one step further, one can consider families of scaling exponents to arrive at a 
single characterizing function for inflationary models. 

The scaling relations between the spectral index and the tensor-to-scalar ratio in dependence of the parameter $\mu$ raise the question 
how this arises in these models. A first scaling relation often considered describes  the variation of the spectral index as a function of the number of e-folds. 
This is usually written as a function that for fixed $p$ only depends on $N$, as in eq. (\ref{nRR-SFIII}).
 However, as shown above in eq. (\ref{nRR-SFII}) and discussed above, this simple universal relation gains its simplicity only in the regime of small $\phi$ 
 and sub-Planckian $\mu$, where it leads to a highly degenerate relation that is not sensitive to the parameter $\mu$ and furthermore
 is not  sensitive to the type of class parameterized by $n$. This form of the $N$-dependence 
 of the spectral index is not implied by the  small-field approximation itself,   i.e. $\phi$ relative to $\mu$. Instead, the spectral index retains both in the 
slow-roll and the small field approximation a dependence on $\mu$ and this leads to the stratification seen in Fig. 2.
Similar issues arise in other types of models considered in the literature, such as Starobinsky inflation \cite{s80, w84, m87}, Higgs inflation \cite{bs07}. % among others.}

%\vfill \eject

In Fig. 3 the energy dependence of the spectral tilt $\delta=1-n_{\cR\cR}$ is considered for the P-ACT-LB-BK18 constraints.
\begin{center}
\includegraphics[scale=0.5]{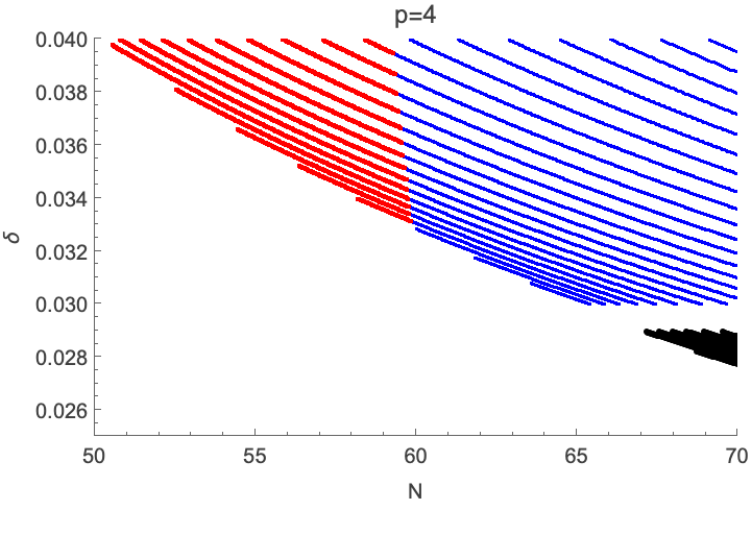}
\includegraphics[scale=0.5]{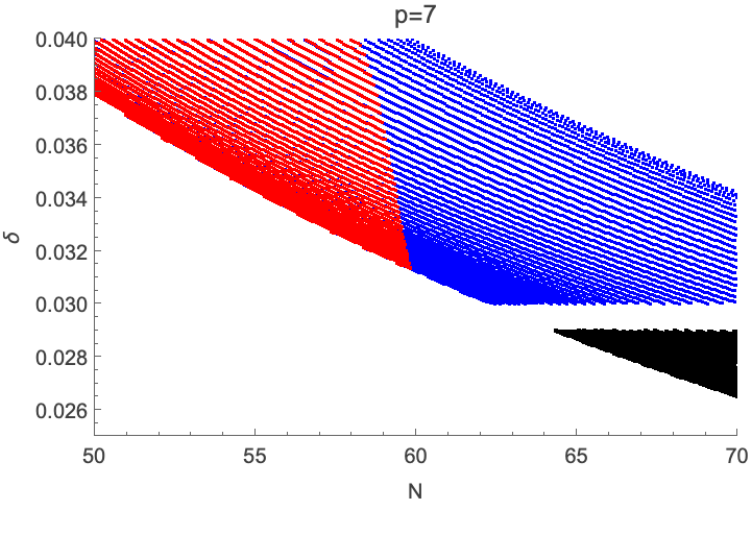}

\includegraphics[scale=0.5]{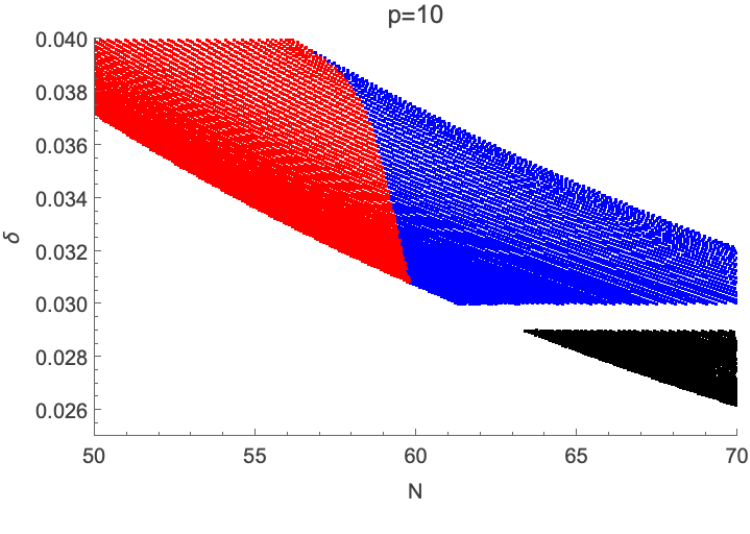} 
\includegraphics[scale=0.5]{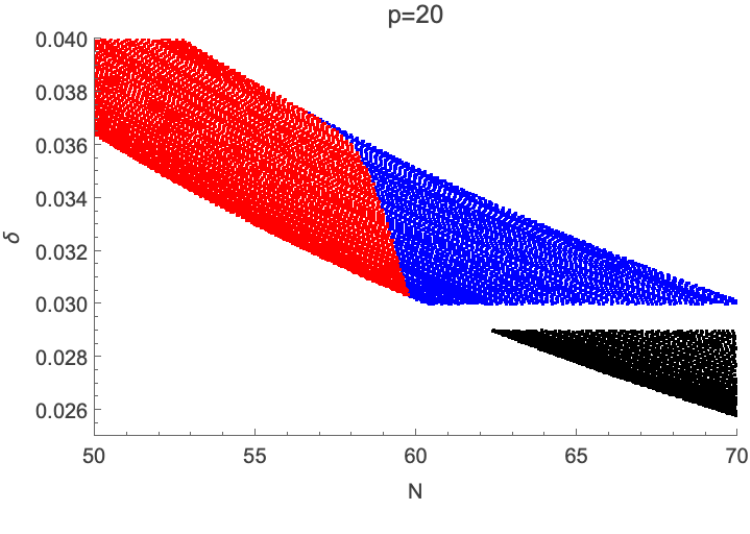} 
\end{center}
\begin{quote}
{\bf Fig. 3}~ The scaling distributions for the spectral tilt $\delta(N)$ for hilltop-squared models with $p=4,7,10, 20$, with $N$ along 
the horizontal axis. The color code here is the same as in Fig. 2.
\end{quote}

The distributions of Fig. 3 show that for all models there exists a large degree of variation for the spectral tilt $\delta(N)$ as $\mu$ runs through the 
allowed parameter range. Instead of a single relation obtained in the approximation of eq. (\ref{nRR-SFIII}), the slow-roll approximation 
leads to the stratifications of Fig. 3 that establishes a significant energy dependence of the spectral tilt $\delta(N, \mu)$ as a 
function of both $N$ and $\mu$.
Each curve for fixed $\mu$ follows a scaling relation which can be viewed as a characteristic of the model. The limit of the energy independent
scaling relation of the approximation $n_{\cR\cR}$ is located at the top of the distribution. This limit is realized for some of the hilltop models
in the {\sc Planck} derived regime, but is absent in the ACT regime for the observables.

The results in Fig. 3 show that models consistent with the observations of the P-ACT-LB-BK18 analysis are all located in the regime where the 
number of e-folds is larger than the value canonically considered in the literature. If one were to restrict the duration of inflation to $N=60$ 
e-folds as the maximal value then none of these 
models would be consistent with the new measurements reported in \cite{act25a}.

Finally, the second standard relation often considered is the scaling behavior of the tensor-to-scalar ratio as a function of the e-folds.
This again has often been considered in a form in which the parameter $\mu$ only enters as an overall scale, as in eq. (\ref{r-SFIII}),
but like in the case of the spectral index, the small-field approximation does lead to an energy dependence as in eq. (\ref{r-SFII}).
In Fig. 4 we present the results for these functions that result from the new observations of the Atacama Cosmology Telescope. 
\begin{center}
\includegraphics[scale=0.5]{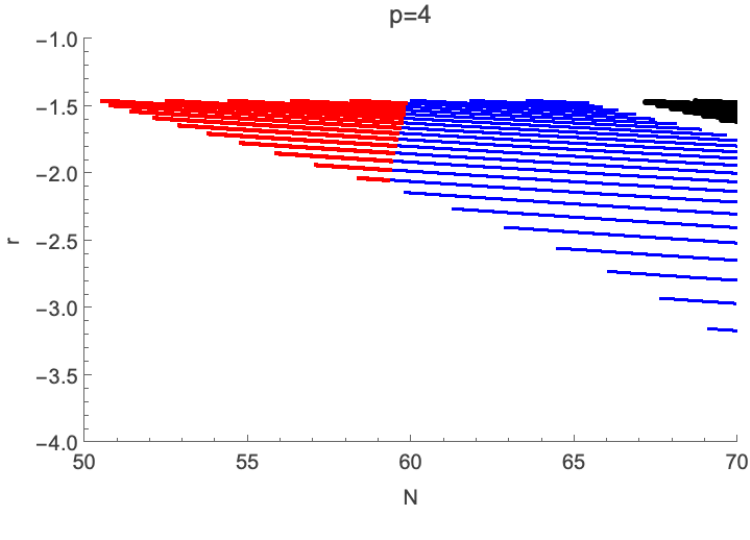}
\includegraphics[scale=0.5]{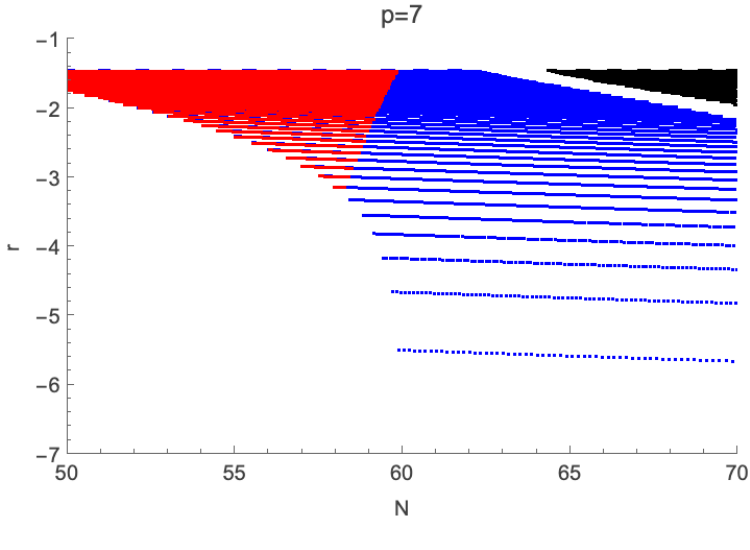}  

\includegraphics[scale=0.5]{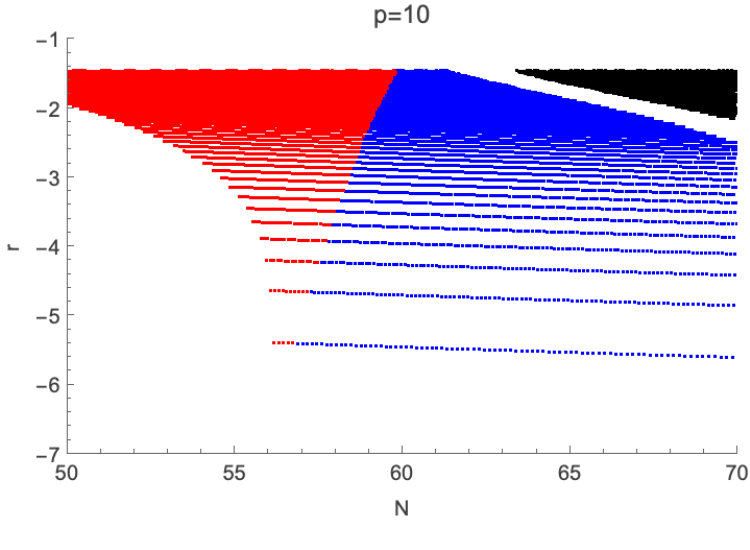} 
\includegraphics[scale=0.5]{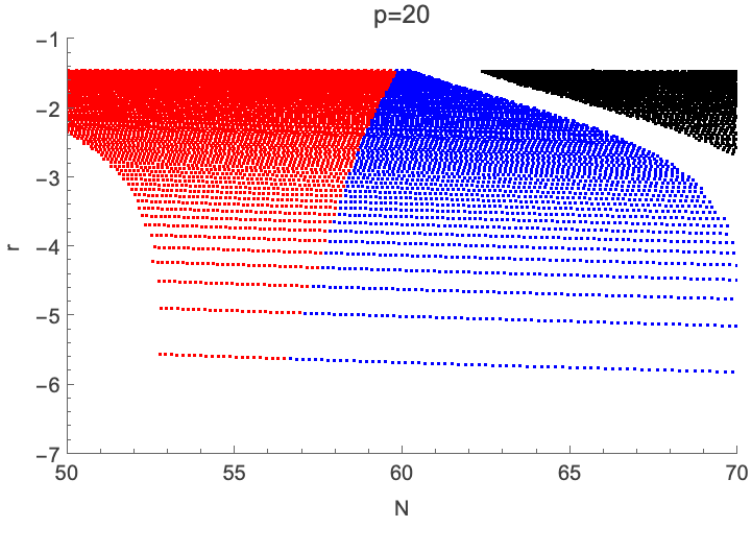} 
\end{center}
\begin{quote}
{\bf Fig. 4}~ The scaling distributions for the function $r(N)$ for hilltop-squared models with $p=4,7,10, 20$. Here the 
number of e-folds $N$ is along the horizontal axis and $r(N)$ is in a log scale along the vertical axis. 
The color code here is the same as in Fig. 2.
\end{quote}

The scaling relations established by the stratification of the results in  Fig. 4, together with the scaling relations in Fig. 3, explain the 
scaling relations in Fig. 2 between the tensor-to-scalar ratio and the spectral index. 
The results in Fig. 4 show that the P-ACT-LB-BK18 regime in parameter space is located toward the larger $\mu$ range,  
for which more stringent future constraints on the tensor-to-scalar  ratio will have significant impact.

We now describe the changes imposed by the ACT result on the hilltop models in comparison with the hilltop-squared models
for the same exponents $p$. It turns out that the structural implications are weaker for the HI$_p$ models than for the HSI$_p$ models.
 While the latter do not, for low $p$, realize the total spectral index range allowed by the ACT data, this range is realized by the hilltop models
considered here. 
A second distinction is that the hilltop class contains for each $p$ realizations for which matter-dominated reheating is viable while 
the class of hilltop-squared models does not. These regions are located in the upper left corner of the ACT domain 
in the $(n_{\cR\cR}, \log~r)$ plane shown in Fig. 5.
One feature that is common to both classes, HSI$_p$ and HI$_p$, is that the viable $\mu$-range shifts higher and no 
sub-Planckian $\mu$ is compatible with ACT. This is different from the {\sc Planck} regime for both classes. 
Furthermore, contrary to the hilltop-squared models, the hilltop class contains for each $p$ regions in parameter space for which the number 
of e-folds lie within the often-considered interval $N_* \in [50, 60]$. It is essentially this parameter region for which 
these realizations admit matter-dominated reheating. We illustrate these differences in Fig. 5 for the hilltop model at $p=4$, 
for which we present the graph in the spectral-tensor plane and the $(N, \delta)$ plane. 
The comparison of the second panel in Fig. 5  with the corresponding panel in Fig. 3 allows us to see explicitly 
not only the dependence of the spectral index on the parameter $\mu$ but also its dependence on the class parameter $n$. This in a way
resolves the doubly degenerate form of the spectral index in its strong small-field approximation given by 
(\ref{nRR-SFIII}) to all parameters $(p, \mu, n)$ of these classes or models.
For higher $p$ the results  are similar in structure, but  with different details, such as a lowering of the minimal tensor ratio and the e-fold number. 
\begin{center}

\includegraphics[scale=0.6]{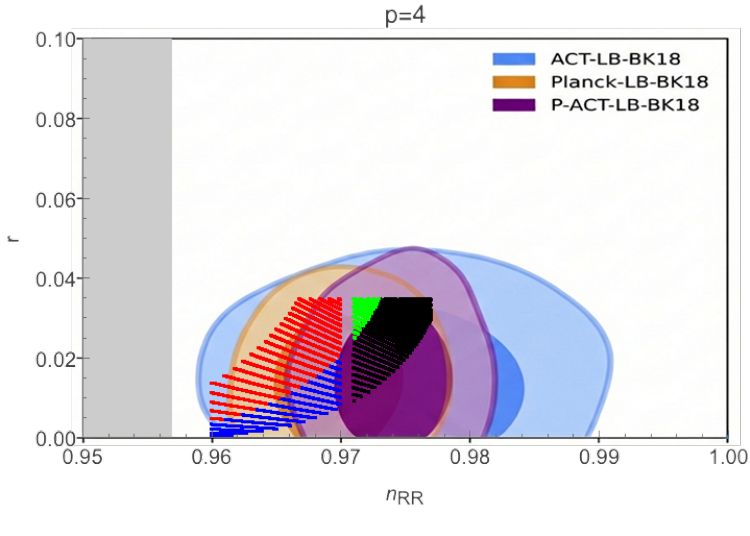}
~~~
\includegraphics[scale=0.6]{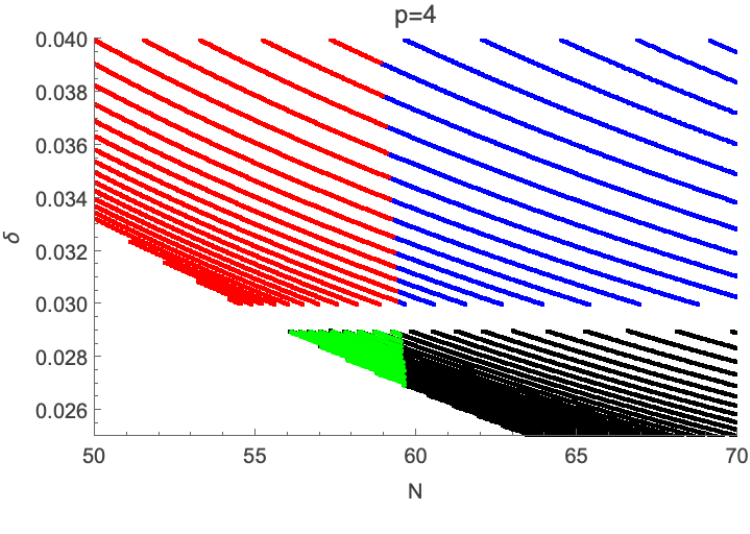}
\end{center}
\begin{quote}
 {\bf Fig. 5} The results for $(n_{\cR\cR}, ~r)$ (left panel) and $(N, \delta)$ (right panel) for the hilltop model at $p=4$.
 The color code is like in Fig. 2 except that green indicates models with matter dominated reheating in the ACT regime.
 The vertical band in the left panel again shows the range of the spectral index in the regime III of small inflaton values and 
 sub-Planckian $\mu$.
 \end{quote}
 
 We finally make a remark about the simplification that occurs in hilltop and hilltop-squared inflation for the super-Planckian 
 parameter space $\mu$ that exists for both the {\sc Planck} and the ACT constraints, and that is preferred by the ACT data. 
 An issue of these hilltop models that was noted early on was that the slow-roll regime can end before inflation ends and hence
 the focus should be on the slow-roll parameter $\eta_V$ instead to $\e_V$. In order to address this issue it is useful to 
 consider the relation (\ref{etaV-vs-eV}) for general $n$. This shows that the relation between the two slow-roll parameters 
 depends in these classes of models on the ratio $\mu/\phi$. If one is interested in the ratio $\eta_V/\e_V$ at the end of inflation
 it is important to have some information about $\mu/\phi_e$. One can check that in the super-Planckian regime the value of 
 $\phi_e$ is close to $\mu$, in which case the equation (\ref{etaV-vs-eV}) reduces in a very good approximation 
 to a numerical proportionality with a factor 
 that depends only on $p$ and $n$, hence the classes parametrized by $n$ behave differently. For the hilltop models the 
 second slow-roll parameter is in this regime much smaller than $\e_V$ while for hilltop-squared inflation they are essentially 
 the same. In a very good approximation the end of inflation can be determined for both types of models with the usual constraint.

\vskip .1truein

\section{Conclusion}

In this paper we have shown that the shift of the spectral index reported by the ACT Collaboration, in combination with external data sets 
from {\sc Planck} and DESI,  has severe consequences for the classes of hilltop and hilltop-squared inflation. 
These implications are more dramatic for hilltop-squared models for several reasons. 
First, the newly determined range for the spectral index can only partially be realized by HSI models, as illustrated in Fig. 2, 
  while in the class of hilltop models each of our models contains realizations that cover the ACT regime of $n_{\cR\cR}$, as shown in Fig. 5.  
The number of e-folds also behaves differently in these two classes of models, where for HSI none of the models lead to 
  realizations for which the number of e-folds  lies in the often-considered range between 50 and 60, while for HI each of the models leads 
  to parameter ranges for which the number of e-folds does take values of 60, or lower.  These results are valid throughout the parameter 
  range of these models and do not rely on the small-field approximation in either its weaker or stronger form.
  We have shown in particular that  in order to gauge the viability of these models it is important to avoid a focus on the regime 
  of small field and sub-Planckian $\mu$ values.
Furthermore, for HSI none of the models leads to matter-dominated reheating for any value of the free parameter $\mu$, while for 
  the hilltop models a parameter region exists for each $p$ for which matter-dominated reheating is viable. 
This parameter range is essentially the range for which the number of e-folds falls in the range between 50 and 60.
As a consequence of the shift to higher $n_{\cR\cR}$ none of the models in either HSI or HI
 admit sub-Planckian energy values for $\mu$. 

Most affected by the new results are the low $p$ models but in all models there is a severe reduction in the viable parameter space. 
While $p=3$  HSI is almost excluded, for $p\geq 4$ there do exist models that satisfy the CMB constraints recently obtained with the Atacama 
Cosmology Telescope for a range of parameter values $\mu$. For $p=3$ HI the constraints are less dramatic and there exists a region of
parameter space where this model is compatible with the P-ACT-LB-BK18 result. 
With the upcoming results of the Simons Observatory \cite{simons18} 
it will be possible to arrive at further constraints that might  exclude the lower $p$ regime of 
these models if gravitational waves are not discovered.

\vskip 0.6truein

\end{document}